\documentclass[prd,twocolumn,floatfix,preprintnumbers,showpacs]{revtex4}
\usepackage{graphics}
\usepackage{longtable}
\usepackage{epsfig}

\begin{document}
\title{Cosmological limit on the neutrino mass}
\author{Steen Hannestad}
\email{steen@nordita.dk}
\affiliation{NORDITA, Blegdamsvej 17, DK-2100 Copenhagen, Denmark}
\date{{\today}}

\begin{abstract}
We have performed a careful analysis of constraints on the neutrino
mass from current cosmological data. Combining data
from the cosmic microwave
background and the 2dF galaxy survey yields an upper limit on the
sum of the three neutrino mass eigenstates of $\sum m_\nu \leq 3$ eV
(95\% conf.), without including additional priors. Including data
from SNIa observations, Big Bang nucleosynthesis, and HST Hubble
key project data on $H_0$ tightens the limit to $\sum m_\nu \leq 2.5$ eV
(95\% conf.). We also perform a Fisher matrix analysis which
illustrates the cosmological parameter degeneracies affecting the
determination of $\sum m_\nu$.
\end{abstract}
\maketitle


\section{introduction}

The absolute value of neutrino masses are very difficult to measure
experimentally. On the other hand, mass differences between neutrino
mass eigenstates, $(m_1,m_2,m_3)$, 
can be measured in neutrino oscillation experiments.
Observations of atmospheric neutrinos suggest a squared mass 
difference of $\delta m^2 \simeq 3 \times 10^{-3}$ eV$^2$
\cite{Fukuda:2000np,Fornengo:2000sr}. While there are still
several viable solutions to the solar neutrino problem the so-called
large mixing angle solution gives by far the best fit with
$\delta m^2 \simeq 5 \times 10^{-5}$ eV$^2$ \cite{sno,Bahcall:2002hv}. 

In the simplest case where neutrino masses are
hierarchical these results suggest that $m_1 \sim 0$, $m_2 \sim 
\delta m_{\rm solar}$, and $m_3 \sim \delta m_{\rm atmospheric}$.
If the hierarchy is inverted 
\cite{Kostelecky:1993dm,Fuller:1995tz,Caldwell:1995vi,Bilenky:1996cb,King:2000ce,He:2002rv}
one instead finds
$m_3 \sim 0$, $m_2 \sim \delta m_{\rm atmospheric}$, and 
$m_1 \sim \delta m_{\rm atmospheric}$.
However, it is also possible that neutrino
masses are degenerate
\cite{Ioannisian:1994nx,Bamert:vc,Mohapatra:1994bg,Minakata:1996vs,Vissani:1997pa,Minakata:1997ja,Ellis:1999my,Casas:1999tp,Casas:1999ac,Ma:1999xq,Adhikari:2000as}, 
$m_1 \sim m_2 \sim m_3 \gg \delta m_{\rm atmospheric}$, 
in which case oscillation experiments are not
useful for determining the absolute mass scale.

Experiments which rely on kinematical effects of the neutrino mass
offer the strongest probe of this overall mass scale. Tritium decay
measurements have been able to put an upper limit on the electron
neutrino mass of 2.2 eV (95\% conf.) \cite{Bonn:tw}.
However, cosmology at present yields an even stronger limit which
is also based on the kinematics of neutrino mass.
Neutrinos decouple at a temperature of 1-2 MeV in the early universe,
shortly before electron-positron annihilation.
Therefore their temperature is lower than the photon temperature
by a factor $(4/11)^{1/3}$. This again means that the total neutrino
number density is related to the photon number density by
\begin{equation}
n_{\nu} = \frac{9}{11} n_\gamma
\end{equation}

Massive neutrinos with masses $m \gg T_0 \sim 2.4 \times 10^{-4}$ eV
are non-relativistic at present and therefore contribute to the
cosmological matter density \cite{Hannestad:1995rs,Dolgov:1997mb,Mangano:2001iu}
\begin{equation}
\Omega_\nu h^2 = \frac{\sum m_\nu}{92.5 \,\, {\rm eV}},
\end{equation}
calculated for a present day photon temperature $T_0 = 2.728$K. Here,
$\sum m_\nu = m_1+m_2+m_3$.
However, because they are so light
these neutrinos free stream on a scale of roughly 
$k \simeq 0.03 m_{\rm eV} \Omega_m^{1/2} \, h \,\, {\rm Mpc}^{-1}$
\cite{dzs,Doroshkevich:tq,Hu:1997mj}. 
Below this scale neutrino perturbations are completely erased and 
therefore the matter power spectrum is suppressed, roughly by
$\Delta P/P \sim -8 \Omega_\nu/\Omega_m$ \cite{Hu:1997mj}.

This power spectrum suppression allows for a determination of the
neutrino mass from measurements of the matter power spectrum on
large scales. This matter spectrum is related to the galaxy correlation
spectrum measured in large scale structure (LSS) surveys via the
bias parameter, $b^2 \equiv P_g(k)/P_m(k)$.
Such analyses have been performed several times before
\cite{Croft:1999mm,Fukugita:1999as}, most recently
using data from the 2dF galaxy survey \cite{Elgaroy:2002bi}. 
This investigation
finds an upper limit of 1.8-2.2 eV for the sum of neutrino masses.
However, this result is based on a relatively limited cosmological
parameter space. 

It should also be noted that, although massive neutrinos have little
impact on the 
cosmic microwave background (CMB)
power spectrum, it is still necessary to include CMB data in any 
analysis in order to determine other cosmological parameters.

In the present paper we perform an extensive analysis, carefully
discussing the issue of parameter degeneracies. The next section
is devoted to a Fisher matrix analysis of the problem which establishes
possible parameter degeneracies and yields a general idea of 
the precision with which the neutrino mass can be measured.
Section III describes a full numerical likelihood analysis of data
from CMB and LSS which yields a robust limit on the neutrino mass.
Finally, section IV contains a discussion and conclusion.


\section{Fisher matrix analysis}

Measuring neutrino masses from cosmological data is quite involved
since for both CMB and LSS the power spectra depend on a plethora
of different parameters in addition to the neutrino mass.
Furthermore, since the CMB and matter power spectra
depend on many different parameters one might
worry that an analysis which is too restricted in parameter space 
could give spuriously strong limits on a given parameter.
Therefore, it is desirable to study possible parameter degeneracies in
a simple way before embarking on a full numerical likelihood analysis.

It is possible to estimate the precision with which the cosmological
model parameters can be extracted from a given hypothetical data set.
The starting point for any parameter extraction is the vector of
data points, $x$. This can be in the form of the raw data, or in
compressed form, typically the power spectrum ($C_l$ for CMB and
$P(k)$ for LSS).

Each data point has contributions from both signal and noise,
$x = x_{\rm signal} + x_{\rm noise}$. If both signal and noise are
Gaussian distributed it is possible to build a likelihood function
from the measured data which has the following form \cite{oh}
\begin{equation}
{\cal L}(\Theta) \propto \exp \left( -\frac{1}{2} x^\dagger 
[C(\Theta)^{-1}] x \right),
\end{equation}
where $\Theta = (\Omega, \Omega_b, H_0, n_s, \tau, \ldots)$ is a vector
describing the given point in model parameter space and 
$C(\Theta) = \langle x x^T \rangle$ 
is the
data covariance matrix.
In the following we shall always work with data in the form of a
set of power spectrum coefficients, $x_i$, which can be either
$C_l$ or $P(k)$.

If the data points are uncorrelated so that the data covariance matrix
is diagonal, the likelihood function can be reduced to
${\cal L} \propto e^{-\chi^2/2}$, where
\begin{equation}
\chi^2 = \sum_{i=1}^{N_{\rm max}} \frac{(x_{i, {\rm obs}}-x_{i,{\rm theory}})^2}
{\sigma(x_i)^2},
\label{eq:chi2}
\end{equation} 
is a $\chi^2$-statistics and $N_{\rm max}$ 
is the number of power spectrum data
points \cite{oh}.

The maximum likelihood is an unbiased estimator, which means that
\begin{equation}
\langle \Theta \rangle = \Theta_0.
\end{equation}
Here $\Theta_0$ indicates the true parameter vector of the underlying
cosmological model and $\langle \Theta \rangle$ is the average estimate
of parameters from maximizing the likelihood function.

The likelihood function should thus peak at $\Theta \simeq \Theta_0$, and
we can expand it to second order around this value.
The first order derivatives are 
zero, and the expression is thus
\begin{widetext}
\begin{equation}
\chi^2  = \chi^2_{\rm min} + \sum_{i,j}(\theta_i-\theta) \left( \sum_{k=1}^{N_{\rm max}}
\frac{1}{\sigma (x_k)^2} \left[\frac{\partial x_k}{\partial \theta_i}
\frac{\partial x_k}{\partial \theta_j} - (x_{k, {\rm obs}}-x_k)
\frac{\partial^2 x_k}{\partial \theta_i \partial \theta_j} \right]\right) (\theta_j-\theta),
\end{equation}
\end{widetext}
where $i,j$ indicate elements in the parameter vector $\Theta$.
The second term in the second derivative can be expected to be very small
because $(x_{k, {\rm obs}}-x_k)$ is in essence just a random measurement error 
which should average out. The remaining term
is usually referred to as the Fisher information matrix
\begin{equation}
F_{ij} = \frac{\partial^2 \chi^2}{\partial \theta_i \partial \theta_j} = 
\sum_{k=1}^{N_{\rm max}}\frac{1}{\sigma (x_k)^2}\frac{\partial x_k}{\partial \theta_i}
\frac{\partial x_k}{\partial \theta_j}.
\label{eq:fisher1}
\end{equation}
The Fisher matrix is closely related to the precision with which the
parameters, $\theta_i$, can be determined.
If all free parameters are to be determined from the data alone without
any priors then it follows from the Cramer-Rao inequality
\cite{kendall} that
\begin{equation}
\sigma(\theta_i) = \sqrt{(F^{-1})_{ii}}
\end{equation}
for an optimal unbiased estimator, such as the maximum likelihood
\cite{tth}.

In order to estimate how degenerate parameter $i$ is with 
another parameter, $j$, one can calculate how $\sigma(\theta_i)$
changes if parameter $j$ is kept fixed instead of free in the
analysis. Starting from the $2 \times 2$ sub-matrix
\begin{equation}
S_{ij} = (F^{-1})_{ij},
\end{equation}
one then finds
\begin{equation}
\sigma_{j \,\, {\rm fixed}}(\theta_i) = \sqrt{\frac{1}{(S^{-1})_{ii}}}
\end{equation}

We therefore define the quantity 
\begin{equation}
r_{ij} = \frac{\sigma_{j \,\, {\rm fixed}}(\theta_i)}{\sigma (\theta_i)}
\leq 1
\label{eq:rij}
\end{equation}
as a measure of the degeneracy between parameters $i$ and $j$.

In order to perform an actual calculation we use the most present
data from CMB and LSS.

{\it CMB data set ---} Several data sets of high precision are now
publicly available.  In addition to the COBE \cite{Bennett:1996ce}
data for small $l$ there are data from BOOMERANG \cite{boom}, MAXIMA
\cite{max}, DASI \cite{dasi} and several other experiments
\cite{WTZ,qmask}.  Wang, Tegmark and Zaldarriaga \cite{WTZ} have
compiled a combined data set from all these available data, including
calibration errors.  In the present work we use this compiled data
set, which is both easy to use and includes all relevant present
information. Altogether there are 24 CMB data points in this compilation.

{\it LSS data set ---} At present, by far the largest survey available
is the 2dF \cite{2dF} of which about 147,000 galaxies have so far been
analysed. Tegmark, Hamilton and Xu \cite{THX} have calculated a power
spectrum, $P(k)$, from this data, which we use in the present work.
The 2dF data extends to very small scales where there are large
effects of non-linearity. Since we only calculate linear power
spectra, we use (in accordance with standard procedure) only data on
scales larger than $k = 0.2 h \,\, {\rm Mpc}^{-1}$, where effects of
non-linearity should be minimal. Making this cut reduces the number
of power spectrum data points to 18.

For calculating the theoretical CMB and matter power spectra we use
the publicly available CMBFAST package \cite{CMBFAST}.
As the set of cosmological parameters we choose
$\Omega_m$, the matter density, $\Omega_k = 1 - \Omega_m - \Omega_\Lambda
-\Omega_\nu$,
the curvature parameter, $\Omega_b$, the baryon density, $H_0$, the
Hubble parameter, $n_s$, the scalar spectral index of the primordial
fluctuation spectrum, $\tau$, the optical depth to reionization,
$Q$, the normalization of the CMB power spectrum, $b$, the 
bias parameter, and $\Omega_\nu$, the neutrino density. 
In all cases we take the number of massive neutrinos to be 3. The reason
is that if neutrinos are to have an impact on CMB and matter spectra
their masses must be much larger than the mass splitting inferred from
atmospheric neutrino observations ($\delta m \sim 0.05-0.1$ eV), and
therefore neutrino masses will be degenerate with $m_1 \sim m_2 \sim m_3
\gg \delta m_{\rm atmospheric}$.

\begin{table*}
\caption{The different priors on parameters other than $\Omega_\nu h^2$ 
used in the analysis.}
\begin{ruledtabular}
\begin{tabular}{lccccccc}
prior type & $\Omega_m$ & $\Omega_b h^2$ & $h$ & $n$ & $\tau$ & $Q$ & $b$ \cr
\colrule
CMB + LSS & 0.1-1 & 0.008 - 0.040 & 0.4-1.0 & 0.66-1.34 & 0-1 & free & free \cr
CMB + LSS + BBN + $H_0$ & 0.1-1 & $0.020 \pm 0.002$ & $0.70 \pm 0.07$ & 0.66-1.34 
& 0-1 & free & free  \cr
CMB + LSS + BBN + $H_0$ + SNIa & $0.28 \pm 0.14$ & $0.020 \pm 0.002$ & $0.70 \pm 0.07$ & 0.66-1.34 
& 0-1 & free & free \cr
\end{tabular}
\end{ruledtabular}
\end{table*}

In principle 
one might include even more parameters in the analysis, such as
$r$, the tensor to scalar ratio of primordial fluctuations. However, $r$
is most likely so close to zero that only future high precision 
experiments may be able to measure it. The same is true for other 
additional parameters. Deviations from the slow-roll prediction of a
simple power-law initial spectrum 
\cite{Hannestad:2000tj,Hannestad:2001nu,griffiths,elgaroy}
or additional relativistic energy
density 
\cite{Jungman:1995bz,Lesgourgues:2000eq,Hannestad:2000hc,Esposito:2000sv,Kneller:2001cd,Hannestad:2001hn,Hansen:2001hi,Bowen:2001in,Dolgov:2002ab}
could be present. However, such effects only appear in cosmological
models which are more complicated than the ``standard'' $\Lambda$CDM model.
The parameters we use fully describe the features of the simplest working
model.

In the end one can check the consistency of the numerical parameter
extraction by calculating the $\chi^2$ per degree of freedom. In Section III
we find that the best fit is consistent with expectations and therefore
it is unlikely that there are other parameters significantly affecting
the power spectra.

Fig.~1 shows the matrix $r_{ij}$, calculated for the WTZ + 2dF
data set errors, around a reference cosmological model with parameters
$\Omega_m = 0.3$, $\Omega_\Lambda = 0.7$, $\Omega_b h^2 =
0.020$, $H_0 = 70 \,\,{\rm km} \, {\rm s}^{-1} \, {\rm Mpc}^{-1}$, $n_s =
1.0$, and $\tau = 0$, i.e.\ the $\Lambda$CDM concordance model.
Note that $r_{ii}$ is an ill-defined quantity. For plotting purposes
we simply put $r_{ii} = 1$, but this has no physical significance.

\begin{figure}[t]
\vspace*{-0.5cm}
\begin{center}
\hspace*{-1.0cm}\epsfysize=10truecm\epsfbox{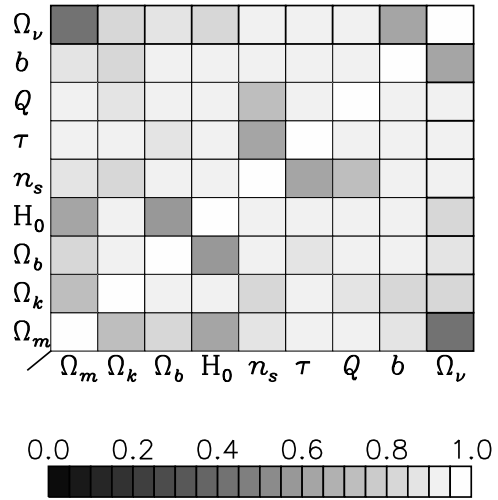}
\end{center}
\vspace*{-2cm}
\caption{Values of the parameter $r_{ij}$, defined in Eq.~(\ref{eq:rij}).}
\label{fig1}
\end{figure}

From this general matrix one can study the $m_\nu$-related degeneracies
more closely. Clearly the two parameters most degenerate with $m_\nu$
are $\Omega_m$, the matter density, and $b$, the bias. 
This is not surprising because
a non-zero neutrino mass has little effect on the CMB acoustic
peaks and a large effect on the matter power spectrum on scales
below the free streaming scale.

Therefore, any parameter which behaves in a similar way will cause degeneracy.
This is indeed the case for $\Omega_m$, if this parameter is changed
while keeping $\Omega_k$ fixed there is little effect on CMB. On the
other hand, changing $\Omega_m$ changes the normalization of the matter
power spectrum at small scales relative to large scales.
Changing $b$ also mimics a non-zero neutrino mass. The reason is that
present day LSS data have large error bars around the free-streaming
scale for light neutrinos in the eV range. On low scales the effect
of massive neutrinos is simply to lower the fluctuation level roughly
as \cite{Hu:1997mj}
\begin{equation}
\frac{\Delta P}{P} \simeq -8 \frac{\Omega_\nu}{\Omega_m},
\end{equation}
i.e.\ it is scale independent and therefore indistinguishable from
changing the bias, $b$. This degeneracy can be broken by precision
measurements around the free-streaming scale where the break in
the power spectrum occurs.
A good example of how the mass limit on neutrinos can be tightened
if bias is fixed comes from Ref.~\cite{Fukugita:1999as}. Here the
mass limit comes from comparing the overall normalization of the spectra
at COBE scales \cite{Bennett:1996ce}
with those on cluster scales \cite{Eke:1996ds}. 
However, we believe that,
at present, keeping bias as a free parameter yields 
a much more robust constraint.
To a much lesser extent the neutrino mass is also degenerate with the
Hubble parameter.

It should be noted that there is little degeneracy with $n_s$, the
spectral index. In Refs.~\cite{Hu:1997mj,Elgaroy:2002bi} a significant
degeneracy between $\Omega_\nu$ and $n_s$ was found when only LSS
data is considered. However, this is broken when CMB data is included
(as is also noted in Ref.~\cite{Hu:1997mj}), the reason being that
changing $n_s$ affects both CMB and matter power spectra, not just
the matter spectrum.

Clearly, it would be desirable to fix the parameters with which the
neutrino mass is most degenerate, $\Omega_m$, $b$, and $H_0$. 
As for $\Omega_m$
one can use the SNIa result $\Omega_m = 0.28 \pm 0.14$ which applies
to a flat universe \cite{Perlmutter:1998np}. 
However, this value is not much more restrictive
than what is found from the CMB+LSS data alone. Fixing the bias
is much more difficult since it is not a physically well understood
parameter. In Elgaroy et al.\ \cite{Elgaroy:2002bi}
bias was kept as a free parameter,
and we follow this line.
$H_0$ has been determined precisely by the HST Hubble key project
to be $H_0 = 70 \pm 7 \,\,{\rm km} \, {\rm s}^{-1} \, {\rm Mpc}^{-1}$
\cite{freedman}. 

If all the above parameters are included in the Fisher matrix analysis
the estimated $1\sigma$ precision on $m_\nu$ is 1.8 eV, equivalent
to a 95\% confidence limit of 3.6 eV.


\section{Numerical results}

The Fisher matrix analysis can only give a general idea of the
constraints which can be found from a given data set. In reality the
likelihood is non-Gaussian and away from the best fit point the
formalism breaks down. In order to get reliable estimates it is
necessary to perform a full numerical likelihood analysis over the
space of cosmological parameters.

In this full numerical likelihood analysis we use a slightly restricted
parameter space with the following free parameters:
$\Omega_m$, $\Omega_b$, $H_0$, $n_s$, $Q$, $b$, and $\tau$. We restrict
the analysis to flat models, $\Omega_k=0$.
This has very little effect
on the analysis because there is little degeneracy between $m_\nu$ and
$\Omega_k$. In order to study the effect of the different priors
we calculate three different cases, the priors for which can be seen
in Table I. The BBN prior on $\Omega_b h^2$ comes from 
Ref.~\cite{Burles:2000zk}. The actual maginalization over parameters
other than $\Omega_\nu h^2$ was performed using a simulated annealing
procedure \cite{Hannestad:wx}.

Fig.~2 shows $\chi^2$ for the three different cases as a function of
the $m_\nu$.
The best fit $\chi^2$ values are 24.81, 25.66, and 25.71 for the
three different priors respectively. In comparison the number of
degrees of freedom are 34, 35, and 36, meaning that the fits are
compatible with expectations, roughly within the 68\% confidence
interval.

\begin{figure}[b]
\vspace*{0.5cm}
\begin{center}
\epsfysize=7truecm\epsfbox{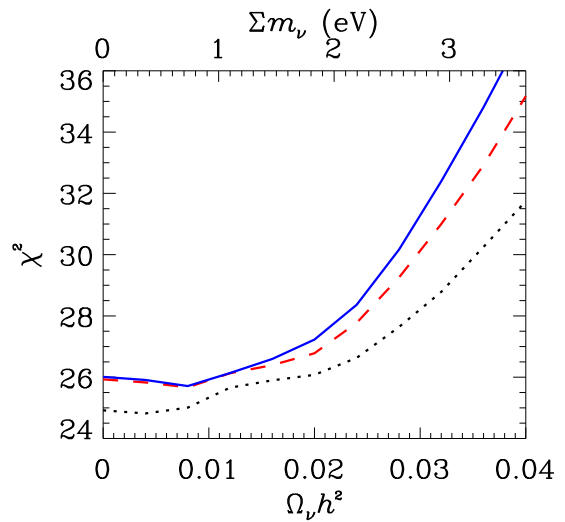}
\end{center}
\vspace*{0cm}
\caption{$\chi^2$ as a function of $\Omega_\nu h^2$, plotted for
the three different priors. The dotted curve is for CMB+LSS, the
dashed for CMB+LSS+BBN+$H_0$, and the full curve for
CMB+LSS+BBN+$H_0$+SNIa.}
\label{fig2}
\end{figure}

We identify the 95\% confidence limit on $m_\nu$ with the point
where $\Delta \chi^2 = 4$. These limits are shown in Table II.
For the most restrictive prior we find a 95\% confidence upper limit
of $\sum m_\nu \leq 2.47$ eV. This is compatible with the findings
of Ref.~\cite{Elgaroy:2002bi} 
who find that $\sum m_\nu \lesssim 1.8-2.2$ eV
for a slightly more restrictive parameter space.

Based on the present analysis we consider 
$\sum m_\nu \leq 3$ eV (95\% conf.) a robust upper limit on the sum
of the neutrino masses. This corresponds roughly to the value found 
for the CMB+LSS data alone without any additional priors.
Even though this value is significantly higher than what is quoted
in Ref.~\cite{Elgaroy:2002bi}, 
it is still much more restrictive than the 
value $\sum m_\nu \leq 4.4$ eV \cite{WTZ} found from CMB and PSCz 
\cite{pscz} data. As is also
discussed in Ref.~\cite{Elgaroy:2002bi} 
the main reason for the improvement is
the much greater precision of the 2dF survey, compared to the PSCz
data \cite{pscz}.

\begin{table*}
\caption{Best fit $\chi^2$ and upper limits on $\sum m_{\nu,{\rm max}}$
for the three different priors.}
\begin{ruledtabular}
\begin{tabular}{lcc}
prior type & best fit $\chi^2$ & $\sum m_{\nu,{\rm max}}$ (eV) (95\% conf.) \cr
\colrule
CMB + LSS & 24.81 & 2.96 \cr
CMB + LSS + BBN + $H_0$ &  25.66 & 2.65 \cr
CMB + LSS + BBN + $H_0$ + SNIa & 25.71 & 2.47 \cr
\end{tabular}
\end{ruledtabular}
\end{table*}


\section{Discussion}

We have studied cosmological constraints on the neutrino masses from
present CMB and LSS data. Initially a Fisher matrix analysis was performed
which illustrates the main degeneracies of $m_\nu$ with other cosmological
parameters used in the analysis. From this simplified analysis it was
estimated that the precision on $\sum m_\nu$ should be roughly 
3.6 eV at 95\% confidence from CMB+LSS data alone.

However, in order to obtain reliable estimates we performed a full 
numerical likelihood analysis. Using reasonable priors on $\Omega_m$,
$\Omega_b h^2$ and $H_0$ we obtained a limit of
$\sum m_\nu \leq 2.47$ eV at 95\% confidence, while using no priors
on the CMB+LSS data yielded $\sum m_\nu \leq 3$ eV, again at 95\%
confidence. We believe this to be a robust upper limit.

Our analysis shows, not surprisingly, that priors are extremely important
for parameter estimation of $m_\nu$. Our most restrictive prior yields
a result similar to that found by Ref.~\cite{Elgaroy:2002bi}, 
while our no-prior 
case yields a significantly looser constraint.

The Fisher matrix analysis showed that the parameter most degenerate with
$m_\nu$ is the bias parameter, $b$. In order to obtain much stronger
limits one must either determine $b$ precisely in an independent way
or obtain better LSS power spectrum statistics around the scale corresponding
to the free-streaming scale for neutrinos, $k \simeq 0.02-0.03 \, h \,\,
{\rm Mpc}^{-1}$. The Sloan Digital Sky Survey \cite{sdss} will measure the
power spectrum shape with higher precision on the relevant scale, and 
this data, combined with CMB data from the MAP experiment
\cite{map}, will 
either push the limit by a factor of at least a few or indeed detect
a non-zero neutrino mass directly.
It was estimated in Ref.~\cite{Hu:1997mj} that $\sum m_\nu \lesssim 0.65$ eV
can be reached.

Finally, we note that the present cosmological limit is significantly
stronger than current laboratory limits. The Mainz tritium experiment
\cite{Bonn:tw}
currently quotes a 95\% upper limit to the $\nu_e$ mass of 2.2 eV, which
translates to a sum of roughly 6.5-7 eV for the three mass eigenstates.
As is also noted in Elgaroy et al.\ \cite{Elgaroy:2002bi}, 
the cosmological limit is 
compatible with the very controversial detection of neutrinoless
double beta decay by the Heidelberg-Moscow experiment
\cite{Klapdor-Kleingrothaus:2001ke}. If this finding
is confirmed it would imply a sum of masses of order 1 eV, within range
of the MAP+SDSS data.

\newcommand\AJ[3]{~Astron. J.{\bf ~#1}, #2~(#3)}
\newcommand\APJ[3]{~Astrophys. J.{\bf ~#1}, #2~ (#3)}
\newcommand\apjl[3]{~Astrophys. J. Lett. {\bf ~#1}, L#2~(#3)}
\newcommand\ass[3]{~Astrophys. Space Sci.{\bf ~#1}, #2~(#3)}
\newcommand\cqg[3]{~Class. Quant. Grav.{\bf ~#1}, #2~(#3)}
\newcommand\mnras[3]{~Mon. Not. R. Astron. Soc.{\bf ~#1}, #2~(#3)}
\newcommand\mpla[3]{~Mod. Phys. Lett. A{\bf ~#1}, #2~(#3)}
\newcommand\npb[3]{~Nucl. Phys. B{\bf ~#1}, #2~(#3)}
\newcommand\plb[3]{~Phys. Lett. B{\bf ~#1}, #2~(#3)}
\newcommand\pr[3]{~Phys. Rev.{\bf ~#1}, #2~(#3)}
\newcommand\PRL[3]{~Phys. Rev. Lett.{\bf ~#1}, #2~(#3)}
\newcommand\PRD[3]{~Phys. Rev. D{\bf ~#1}, #2~(#3)}
\newcommand\prog[3]{~Prog. Theor. Phys.{\bf ~#1}, #2~(#3)}
\newcommand\RMP[3]{~Rev. Mod. Phys.{\bf ~#1}, #2~(#3)}

\end{document}